\documentclass[
  reprint,
  superscriptaddress,
  amsmath,amssymb,
  aps,
  prb,
]{revtex4-2}
\usepackage{comment}
\usepackage{ragged2e}
\usepackage{caption}
\usepackage{graphicx}
\usepackage{dcolumn}
\usepackage{bm}
\usepackage{physics}
\usepackage{hyperref}
\usepackage[dvipsnames]{xcolor}
\usepackage{subcaption}
\usepackage{svg}
\usepackage{verbatim}

\usepackage{orcidlink}
\newcommand{\orciddaniel}{\orcidlink{0000-0001-7658-3546}}
\newcommand{\orcidsimone}{\orcidlink{0000-0002-8882-2169}}
\newcommand{\orcidmarco}{\orcidlink{0000-0002-3574-7003}}
\newcommand{\orcidsimonen}{\orcidlink{0000-0001-8364-6330}}
\newcommand{\orcidluka}{\orcidlink{0000-0002-9800-9200}}
\newcommand{\orcidmarcor}{\orcidlink{0000-0002-4544-3513}}
\newcommand{\orcidpietro}{\orcidlink{0000-0001-5279-7064}}

\definecolor{smoothred}{HTML}{C5232F}

\RequirePackage{fix-cm}

\newcommand{\simone}[1]{{\color{black} #1}}
\begin{document}

\title{Boundary-induced geometrical frustration in a square Rydberg-atom lattice}

\author{Marco Rigobello\orcidmarcor}
\altaffiliation{These authors contributed equally.}
\affiliation{Max-Planck-Institut für Quantenoptik, Hans-Kopfermann-Str.~1, D-85748 Garching, Germany}

\author{Luka Pavešić\orcidluka}
\altaffiliation{These authors contributed equally.}
\affiliation{Dipartimento di Fisica e Astronomia ``Galileo Galilei'', Università degli Studi di Padova, 35131, Padova, Italy}
\affiliation{Istituto Nazionale di Fisica Nucleare, Sezione di Padova, Padova, Italy}

\author{Daniel Jaschke\orciddaniel}
\affiliation{Planqc,  Münchener Str. 34, 85748 Garching, Germany}

\author{Pietro Silvi\orcidpietro}
\affiliation{Dipartimento di Fisica e Astronomia ``Galileo Galilei'', Università degli Studi di Padova, 35131, Padova, Italy}
\affiliation{Istituto Nazionale di Fisica Nucleare, Sezione di Padova, Padova, Italy}
\affiliation{Padua Quantum Technologies Research Center, Università degli Studi di Padova, Italy}

\author{Marco~Di~Liberto\orcidmarco}
\affiliation{Dipartimento di Fisica e Astronomia ``Galileo Galilei'', Università degli Studi di Padova, 35131, Padova, Italy}
\affiliation{Istituto Nazionale di Fisica Nucleare, Sezione di Padova, Padova, Italy}
\affiliation{Padua Quantum Technologies Research Center, Università degli Studi di Padova, Italy}

\author{Simone Montangero\orcidsimone}
\affiliation{Dipartimento di Fisica e Astronomia ``Galileo Galilei'', Università degli Studi di Padova, 35131, Padova, Italy}
\affiliation{Istituto Nazionale di Fisica Nucleare, Sezione di Padova, Padova, Italy}
\affiliation{Padua Quantum Technologies Research Center, Università degli Studi di Padova, Italy}

\author{Simone Notarnicola\orcidsimonen}
\email{simone.notarnicola@unipd.it}
\affiliation{Dipartimento di Fisica e Astronomia ``Galileo Galilei'', Università degli Studi di Padova, 35131, Padova, Italy}
\affiliation{Istituto Nazionale di Fisica Nucleare, Sezione di Padova, Padova, Italy}
\date{\today}

\begin{abstract}
We study geometric defects in square-lattice Rydberg-atom arrays induced by lattice-size parity in the striated density-wave phase.
While odd-sized lattices select a single striated domain, even-sized lattices seed incompatible domains at the four corners, resulting in a $\pi/2$-rotation invariant state with a geometrical defect emanating from the lattice center.
At larger detunings, we instead identify a $\pi$-rotation-invariant state with two domains nucleated by opposite corners and separated by a one-dimensional diagonal defect. 
We probe the state-preparation protocol on the Aquila atomic quantum processor and benchmark it with equilibrium and real-time dynamics tensor-network simulations. 
Our results assess the role of boundaries and lattice size in determining the global ordering in programmable quantum matter.
\end{abstract}

\maketitle

\section{Introduction}

\begin{figure}[t!]
\centering
    \includegraphics[width=1.0\linewidth]{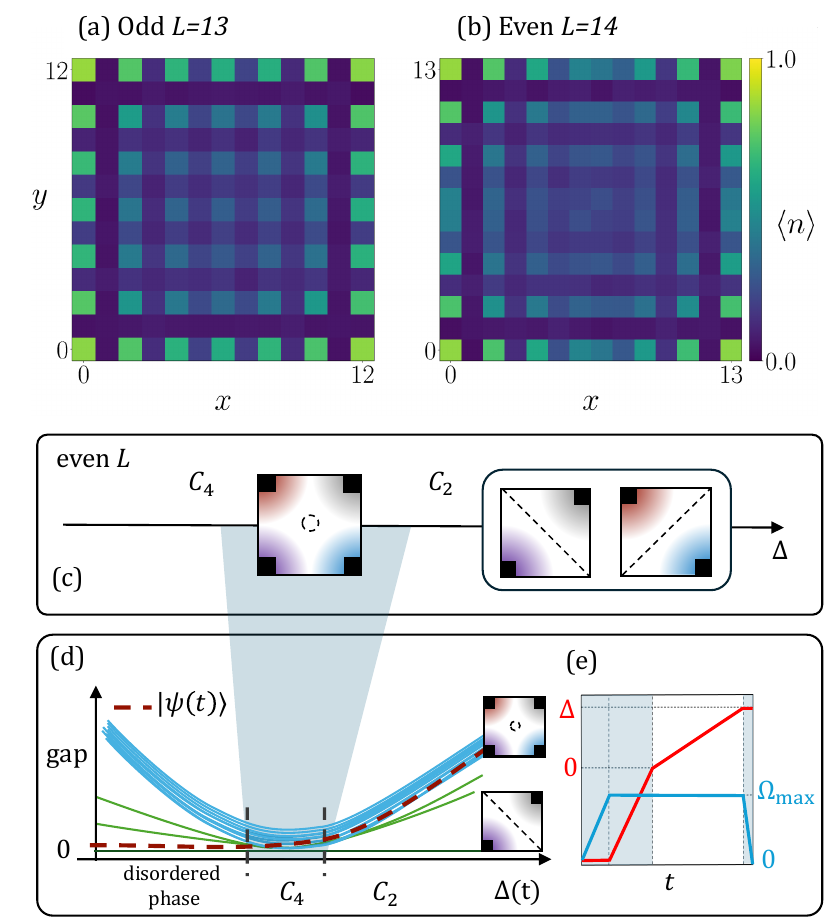}
    \caption{\justifying
(a)--(b) Experimental Rydberg excitation density $\langle n\rangle$ measured after the ramp at final detuning $\Delta=2.0$. For the odd-sized array ($L=13$), the boundary conditions are compatible with a single striated domain. For the even-sized array ($L=14$), the four corners seed incompatible striated patterns, generating a geometrical defect in the bulk. 
(c) Boundary-induced frustrated regimes for even $L$. As the detuning $\Delta$ is increased, the system enters a globally $\mathcal{C}_4$-symmetric frustrated regime, with four striated domains related by $\pi/2$ rotations. At larger $\Delta$, the ground state becomes the superposition of two $\mathcal{C}_2$-invariant configurations with a one-dimensional diagonal defect.
(d) Schematic spectrum along the ramp. 
Green and blue lines represent the gap of instantaneous \simone{eigenstates at a given $\Delta$. We argue for the existence of a dense set of states (blue lines) that preserves the global $\mathcal{C}_4$ rotational structure.} The darkest, bottom lines represent the ground state. 
During the non-equilibrium dynamics, the evolved state $|\psi(t)\rangle$ remains in the $\mathcal{C}_4$-symmetric configuration \simone{(dashed brown line)}, while the ground state transforms into a superposition of the two  $\mathcal{C}_2$-symmetric configurations.
(e) Experimental ramp protocol. \simone{The detuning $\Delta(t)$ goes from negative to zero and then to the final value $\Delta$, while the Rabi frequency is set to the maximum value $\Omega_\mathrm{max}$ allowed by Aquila.} The white central region indicates the portion of the protocol where the detuning crosses the transition into the striated phase and the $\mathcal{C}_4$ and $\mathcal{C}_2$ regimes emerge. Details of the experimental protocol are reported in Appendix~\ref{app:experimental_details}.
}
\label{fig:FIG1}
\end{figure}

Reconfigurable Rydberg-atom arrays are among the most versatile platforms for quantum simulation~\cite{Bloch2008ManyBodySystems, Georgescu2014QuantumSimulation, Fauseweh2024QuantumManyBodySimulations,Greiner2002SuperfluidMott,Schneider2008FermiHubbard,Jotzu2014HaldaneModel,Roushan2014TopologicalTransitions,Labuhn2016RydbergIsing,Turner2018ManyBodyScars,Bernien2017, Browaeys2020}. They can realize large lattices with programmable geometries, which has led to the observation of spin-density-wave phases~\cite{Scholl2021RydbergAntiferromagnets,Ebadi2021QuantumPhases,PhysRevLett.124.103601} and topological states~\cite{semeghini_topological}. Strong van der Waals interactions, together with local control, have also enabled the study of nonthermal dynamics~\cite{bluvsteinscars2021, manovitz_quantum_2025} and the engineering of non-native interactions through Floquet and Rydberg-dressing protocols~\cite{5qhh-322q,geim2026engineeringquantumcriticalitydynamics,dressedweckesser}.
In these experiments, the system size and geometry are typically chosen to be commensurate with the spatial periodicities of the phases under investigation to minimize boundary effects~\cite{Ebadi2021QuantumPhases}. 
However, this does not always tell the whole story.

Already in one-dimensional arrays, the interplay between lattice parity and long-range Rydberg interactions may favor symmetry-broken configurations over a coherent superposition of the two \(\mathbb{Z}_2\) ordered states. In such cases, optimal control approaches are required to restore coherent macroscopic superpositions~\cite{Omran2019CatStates}.
\simone{It has also recently been shown that changing the lattice parity in a one-dimensional or a ladder geometry can change the nature of phase transitions and generate multipartite entangled states~\cite{maric_quantum_2020,catalano2025new,catalano2026experimentalpreparationwstates,catalano2026topologicalfrustrationquantumresources}.}

Here, we study the role of the parity of the lattice size in two dimensions.
Square Rydberg-atom arrays continue to attract attention for their relevance to floating phases, lattice defects, finite-temperature effects, and distinct bulk and boundary ordering transitions~\cite{
zhang_probing_2025,
Wang_2025,
han2025exploringfinitetemperaturebehaviorrydberg,
bulk_boundary_kalinowski_2022}. They also serve as a valuable testbed for novel numerical methods~\cite{PhysRevResearch.5.013026}. 
Nevertheless, the potential of boundary nucleation as a tool to control bulk ordering and engineer frustration has remained largely unexplored.
\simone{In this work, using the neutral-atom quantum processor Aquila~\cite{wurtz2023aquilaqueras256qubitneutralatom, lu_digitalanalog_2024, Gonz_lez_Cuadra_2025}, we leverage lattice parity and boundary effects to engineer frustration and defect structures in programmable quantum matter, without modifying the local connectivity of the model.
To benchmark our quantum simulation results, we use equilibrium (DMRG) and out-of-equilibrium (TDVP) simulations with Tree Tensor Networks (TTN)~\cite{Tagliacozzo2009, Murg2010, felser_introduction_2026}.
See App.~\ref{app:ttns} for details.
}

\simone{We drive an $L \times L$ lattice of Rydberg atoms into the striated density-wave phase~\cite{Samajdar_2020,Ebadi2021QuantumPhases,bulk_boundary_kalinowski_2022,ORourke2023} 
 (see App.~\ref{app:extr_QPT_odd} for details), using the linear ramp sketched in Fig.~\ref{fig:FIG1}(e).
}
We find that the system's behavior strongly depends on the parity of the lattice size. The contrast between odd and even lattices is immediately apparent from the distribution of the Rydberg excitation density in Fig.~\ref{fig:FIG1}. For odd $L$ (Fig.~\ref{fig:FIG1}(a)),
the lattice is commensurate with the striated ordering, and the ramp produces a single striated domain extending across the whole array\simone{, with the atoms at its four corners in their Rydberg excited state.}
For even $L$ (Fig.~\ref{fig:FIG1}(b)), instead, Rydberg excitations form at the four corners, seeding competing striated orders. This leads to domain walls extending from the boundary and throughout the bulk, and the appearance of a central geometrical defect at the four-domain junction point.

In numerical simulations of the ground state, increasing the detuning $\Delta$ further reveals a transition into another distinct regime, also only present in even-$L$ lattices, where the system favors two striated domains seeded by opposite corners.
Accordingly, the \simone{geometry} of the defect changes \simone{from a central defect to an extended diagonal interface}. This order is sketched in Fig.~\ref{fig:FIG1}(c).

Interestingly, the experimental linear-ramp protocol does not reach this state.
As we increase the detuning, the system remains trapped in the manifold of the four corner-seed domains, as sketched in Fig.~\ref{fig:FIG1}(d).

The rest of the paper is organized as follows: Sec.~\ref {sec:model} introduces the Hamiltonian governing the system. Sec.~\ref{sec:central_defect} discusses the four-domain $\mathcal{C}_4$-symmetric ground state, while in Sec.~\ref{sec:diagonal_defect} we discuss how this breaks into $\mathcal{C}_2$-symmetric diagonal-interface configurations. In Sec.~\ref{sec:conclusions} we summarize our results and present possible outlooks.

\section{Model and protocol}
\label{sec:model}

We consider an $L\times L$ square lattice of ${}^{87}$Rb atoms, described by the effective spin-$1/2$ Rydberg Hamiltonian
\begin{equation}
H(t)
=
\sum_{\mathbf{r}}
\left[
\frac{\Omega(t)}{2}\sigma^x_{\mathbf{r}}
-
\Delta(t)n_{\mathbf{r}}
\right]
+
\frac{1}{2}
\sum_{\mathbf{r}\neq\mathbf{r}'}
V_{\mathbf{r}\mathbf{r}'}
n_{\mathbf{r}}n_{\mathbf{r}'},
\end{equation}
where $n_\mathbf{r}=(1+\sigma^z_\mathbf{r})/2$ is the Rydberg excitation number operator on lattice site $\mathbf{r}$, $\Omega(t)$ is the Rabi frequency, $\Delta(t)$ is the detuning, and $
V_{\mathbf{r}\mathbf{r}'}
=
\frac{C_6}{|\mathbf{r}-\mathbf{r}'|^6}
$. For the Rydberg state $|r\rangle = |70\,\mathrm{nS}\rangle$ used in the Aquila processor, $C_6 = 2\pi\times 862690 \,\mathrm{MHz}\,\mu m^6$. 
In the protocol, we set the plateau value of the Rabi frequency to the maximum allowed by the quantum processor, $\Omega_{\rm max} = 15.8\,\mathrm{rad/\mu s} \approx 2\pi \times 2.5\,\mathrm{MHz}$. We express all energy scales in units of $\Omega_{\rm max}$, so that the nearest-neighbor interaction reads
$
V_1
=
\frac{C_6}{a^6\Omega_{\rm max}}
=
\left(\frac{R_b}{a}\right)^6,
$
where $a$ is the lattice spacing.
In numerical simulations, we keep all interaction terms within $|\mathbf{r}-\mathbf{r}'|\leq 2\sqrt{2}\,a$.
 
We set the interatomic spacing such that $R_b/a = 1.46$, extending the Rydberg blockade beyond nearest neighbors and thus favoring the striated order~\cite{Ebadi2021QuantumPhases}.
Then, we tune the Hamiltonian parameters to drive the system from the disordered phase to a final detuning $\Delta$, where we expect the ground state to be in the striated phase. The protocol is sketched in Fig.~\ref{fig:FIG1}(e) (see App.~\ref{app:experimental_details} for details).
We perform experiments on $L=13$ and $L=14$ lattices, i.e., the largest sizes at which we can probe the striated phase using the Aquila processor.

\section{Central geometrical defect}
\label{sec:central_defect}

During the ramp, at detuning values above the boundary critical value~\cite{bulk_boundary_kalinowski_2022}, the corners nucleate the ordered phase. In odd-sized lattices, the four corners share the same striated pattern and generate a single domain throughout the bulk. This dynamical feature is also reflected in the equilibrium properties, where the boundary orders before the bulk~\cite{bulk_boundary_kalinowski_2022}. In Appendix~\ref{app:extr_QPT_odd}, we show that the subsequent bulk ordering is compatible with a disordered-to-striated extraordinary second-order phase transition.
Conversely, in even-sized lattices, the corners belong to different striated patterns.
When the excitation of the corner sites becomes energetically favored, they seed four competing domains which cannot be extended simultaneously throughout the entire lattice. 

\begin{figure}[t!]
\centering
    \includegraphics[width=1.0\linewidth]{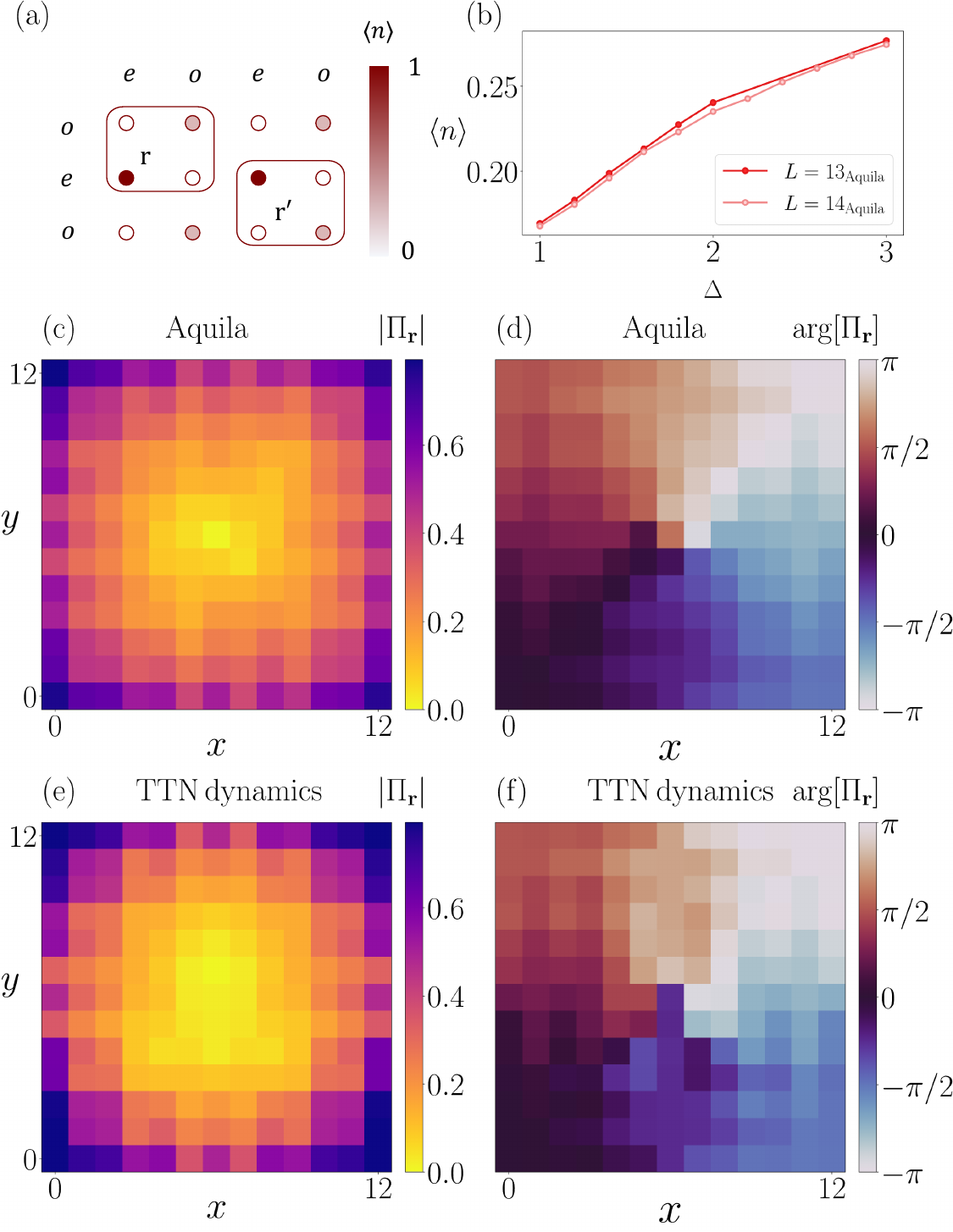}
    \caption{\justifying
(a) Computation of $\Pi_{\mathbf{r}}$ on a plaquette, where $\mathbf{r}$ is the bottom-left site. The parameter $\Pi_{\mathbf{r}}$ depends on the four local densities in the plaquette, taken according to the parity of their coordinates.
(b) Experimentally measured Rydberg excitation density $\langle n\rangle$ as a function of the final detuning $\Delta$. The agreement between the results for $L=13$ and $L=14$ shows that the total density does not depend strongly on the parity of $L$.
(c)--(f) Modulus and phase of $\Pi_{\mathbf{r}}$ for $L=14$ at $\Delta=2.0$, obtained from the experimental data (c,d) and with TTN dynamics (e,f). The agreement between the two shows that the ramp dynamics captures the formation of the four boundary domains and the suppression of local striated order in the bulk.
}
\label{fig:FIG3_compl_pi_tot}
\end{figure}

To measure the striated order and distinguish between the four possible patterns, we define a complex plaquette order parameter $\Pi_\mathbf{r}$, constructed out of the Rydberg excitation density. 
In each $(2\times2)$-site plaquette, we label the sites according to their parity, and define
\begin{equation}\label{eq:plaquette_ord_par}
\Pi_\mathbf{r}=
n_{ee}-n_{oo} + i
(n_{eo}-n_{oe}),
\end{equation}
where $\mathbf{r}$ is the coordinate of the bottom-left site of the plaquette, and $n_{p_xp_y}$ indicates the Rydberg density of the site $(x,y)$ whose coordinates have parity $(p_x,p_y)$. \simone{On the four possible striated domains, $\Pi_\mathbf{r} = i^{k}(1-\eta)$, with $k\in\{0,\dots,3\}$ and where $(1-\eta) > 0$ is the density difference between the nonempty sites in the corresponding plaquette.}
All plaquettes belonging to the same striated domain have the same value of $\Pi_\mathbf{r}$, as one can verify for $\Pi_\mathbf{r}$ and $\Pi_{\mathbf{r}'}$ in Fig.~\ref{fig:FIG3_compl_pi_tot}(a).

In Fig.~\ref{fig:FIG3_compl_pi_tot}(c),(d), we plot the modulus and phase of $\Pi_\mathbf{r}$ for the density profile in Fig.~\ref{fig:FIG1}(b), corresponding to a $\Delta=2.0$ state at the end of the experimental protocol. 
We find qualitative agreement with the results from TTN dynamics, shown in Fig.~\ref{fig:FIG3_compl_pi_tot}(e),(f). 
\simone{In agreement with the density profile, $|\Pi_\mathbf{r}|$ is larger around the corners, where excitations are localized and form striated domains, while the incompatibility between the domains is reflected in $|\Pi_\mathbf{r}|\rightarrow0$ as shown in Fig.~\ref{fig:FIG3_compl_pi_tot}(c), indicating the presence of a geometrical defect in the bulk.

The suppression of $|\Pi_\mathbf{r}|$ in the bulk, however, does not mean that the bulk is depleted of Rydberg excitations: Fig.~\ref{fig:FIG3_compl_pi_tot}(b), shows that the total excitation density in even-sized lattices grows similarly to that of odd-sized ones, in which the excitations form a single, densely packed striated domain. 
Therefore, also in even-sized lattices, the bulk enters a densely packed regime, but without a unique striated pattern.
The local suppression of the striated order we observe in TDVP numerical simulations is thus due to the superposition of 
competing microscopic configurations with differently shaped domains that prevent the selection of a unique ordered pattern. The experiment exhibits the same configurational averaging, although our density measurements do not allow us to assess coherence. 
}

\begin{figure}[t!]
\centering
    \includegraphics[width=1\linewidth]{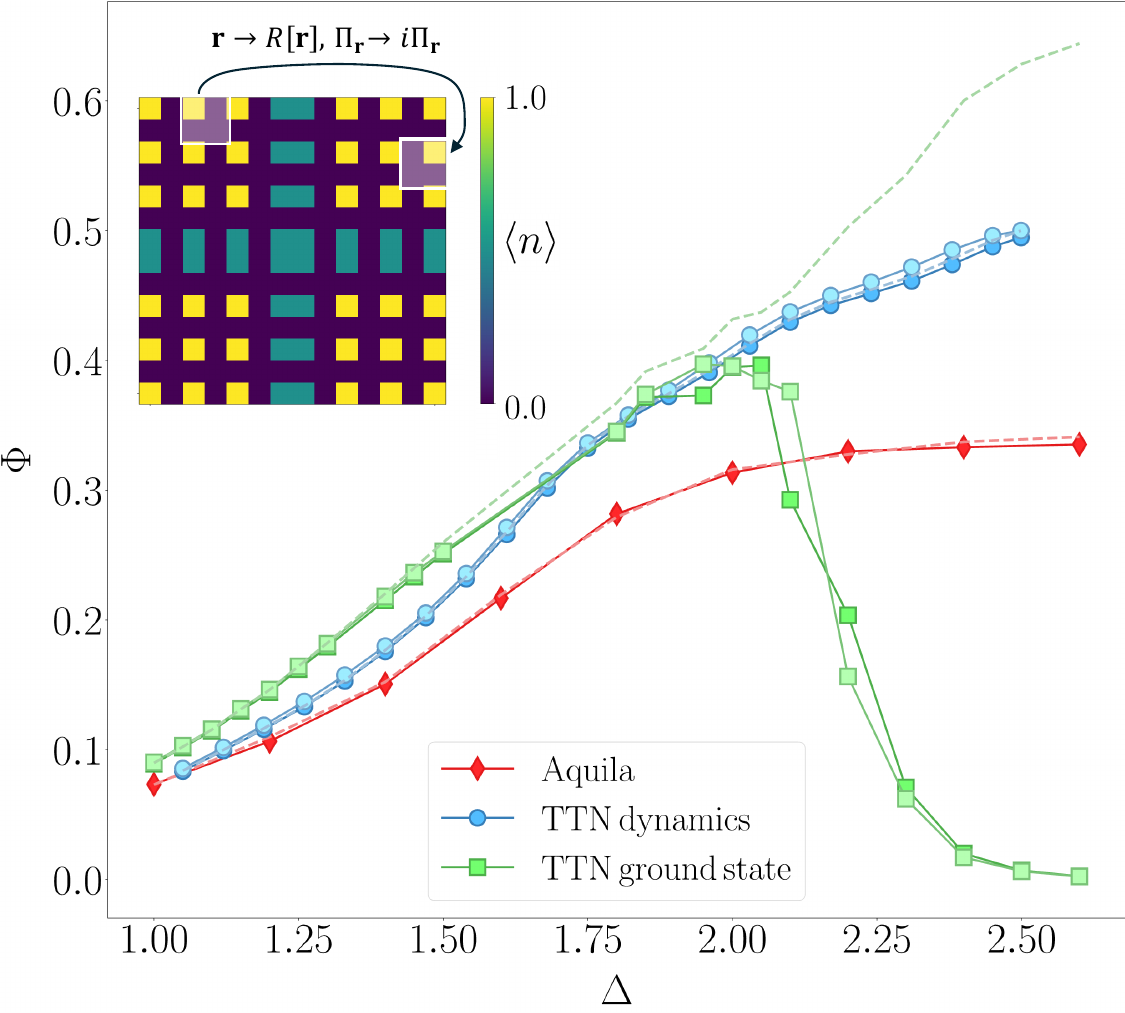}
    \caption{\justifying  
    Rotational correlator $\Phi$ computed for $L=14$ as a function of the final detuning $\Delta$.
    We compare the experimental results obtained with Aquila (red diamonds), real-time TTN dynamics (blue circles), and TTN ground-state calculations (green squares).
    The dashed lines show $\Pi^2$ for each case in the same color.
    For the experimental data, we estimated statistical errors via bootstrap sampling, obtaining errors $\sim 10^{-3}$.
    For numerical simulations, darker shades correspond to larger bond dimensions. 
    For the dynamical simulations, different shades correspond to bond dimensions $m=100,150$, while for the ground-state calculations they correspond to $m=200,250$.
    For the ground-state simulations, we find relative convergence in energy to $10^{-3}$. 
    Inset: Sketch of a $\mathcal{C}_4$-symmetric \simone{state} which maximizes $\Phi$. In this case, $\Phi = \Pi^2$.
}
    \label{fig:FIG4_rot_inv_ord_par_even_exp_dyn_ttn}
\end{figure}

The phase of $\Pi_\mathbf{r}$, Fig.~\ref{fig:FIG3_compl_pi_tot}(d,f), smoothly interpolates between $-\pi$ and $\pi$ as it winds around the central defect, compatibly with the simultaneous presence of all four striated orders in the lattice.
The state thus combines local striated order, visible in each corner domain, with a globally $\mathcal{C}_4$-symmetric arrangement. 
To verify this, we define a global order parameter for the $\mathcal{C}_4$ regime: the boundary rotational correlator,
\begin{equation}\label{eq:rot_inv_plaq_corr}
\Phi
=
\frac{1}{|\partial\Lambda_p|}
\left|
\sum_{\mathbf{r}\in\partial\Lambda_p}
\left\langle \mathrm{Im}\left[
 \Pi_\mathbf{r}^\dagger\,
\Pi_{R[\mathbf{r}]} \right]
\right\rangle
\right|,
\end{equation}
where $\Lambda_p$ denotes the lattice of $(L-1)\times(L-1)$ plaquettes and the average is taken over its boundary, while $R$ is a spatial $\pi/2$ rotation of $\Lambda_p$ around its center:\begin{equation}\label{eq:R_operator}
\mathbf{r}
=
\begin{pmatrix}
x\\
y
\end{pmatrix}
\;\to\;
R[\mathbf{r}]
=
\begin{pmatrix}
y\\
L-2-x
\end{pmatrix},
\end{equation}
with $x,y\in\{0,\dots,L-2\}$.

We first estimate $\Phi$ on a reference $\mathcal{C}_4$-symmetric state, like the cartoon state in the inset of Fig.~\ref{fig:FIG4_rot_inv_ord_par_even_exp_dyn_ttn}, which consists of four striated domains separated by domain walls localized along the horizontal and vertical axes.
Here, the plaquette order at a boundary point $\mathbf{r}$, after the internal transformation $i\Pi_\mathbf{r}$, matches the same pattern at the spatially rotated point $R[\mathbf{r}]$, namely $\left\langle\Pi_\mathbf{R[x]} \right\rangle= i\left\langle \Pi_\mathbf{r}\right\rangle$. In this case, therefore, we have 
$\left\langle \mathrm{Im}\left[\Pi_\mathbf{r}^\dagger\,\Pi_{R[\mathbf{r}]} \right]\right\rangle  = |\langle\Pi_\mathbf{r}\rangle|^2$ for each boundary point and $\Phi = \sum_{\mathbf{r}\in\partial\Lambda_p}{\left|\langle\Pi_\mathbf{r}\rangle\right|^2} = :\Pi^2$.
If the state 
consisted of a \simone{single} non-$\mathcal{C}_4$-invariant configuration with four striated domains, we would have instead $\Phi/\Pi^2 <1$.

In Fig.~\ref{fig:FIG4_rot_inv_ord_par_even_exp_dyn_ttn}, we plot $\Phi$ (continuous lines) and $\Pi^2$ (dashed lines) as a function of $\Delta$. We compare the experimental results (red diamonds), the TTN state preparation obtained with TDVP (blue circles), and the TTN ground state obtained with DMRG (green squares). The latter indicates the value of $\Phi$ in the instantaneous ground state at a given $\Delta$. 
Showing $\Phi$ and $\Pi^2$ allows us to infer both the strength of the striated order along the boundary ($\Pi^2$) and its $\mathcal{C}_4$ invariance ($\Phi$).

At values of $\Delta\lesssim 2.0$, the experimental and numerical curves display the same qualitative behavior: $\Phi$ increases upon entering the striated regime as the four boundary domains form.
For all cases, we observe $\Phi\approx\Pi^2$, consistent with the $\mathcal{C}_4$ invariance of the equilibrium and out-of-equilibrium states.
Unlike the state sketched in Fig.~\ref{fig:FIG4_rot_inv_ord_par_even_exp_dyn_ttn}, in the states we measure and prepare experimentally, domain walls can move along the lattice boundaries.
Despite exhibiting global $\mathcal{C}_4$-invariance, such states contain non-symmetric microscopic configurations that lower $\Phi$ with respect to the reference state.

\simone{
The precise value of $\Phi$ should nevertheless be interpreted with some care. In the experiment, finite coherence time, atom loss, state-preparation imperfections, and readout errors can reduce the measured correlations. In tensor-network simulations, finite bond dimension limits 
the correlations that can be faithfully captured and can affect configurational averaging over domain-wall positions.
More details on this are in Appendix~\ref{app:gs_symmetry_selection}.

At larger detuning, $\Delta\gtrsim2.0$, the value of $\Phi$ in the ground state drops rapidly, signaling a change in the structure of the ground state from the $\mathcal{C}_4$-symmetric frustrated regime. We discuss this next.}

\section{Diagonal geometrical defect}
\label{sec:diagonal_defect}
\simone{
At larger detuning, $\Delta\gtrsim2.0$, the $\mathcal{C}_4$-symmetric ground state gives way to a superposition of two $\mathcal{C}_2$-symmetric states. In each of them, two opposite corners are excited, while the other two stay unexcited. The excited corners seed two striated domains separated by a diagonal interface.}
\begin{figure}[t!]
\centering
    \includegraphics[width=1.0\linewidth]{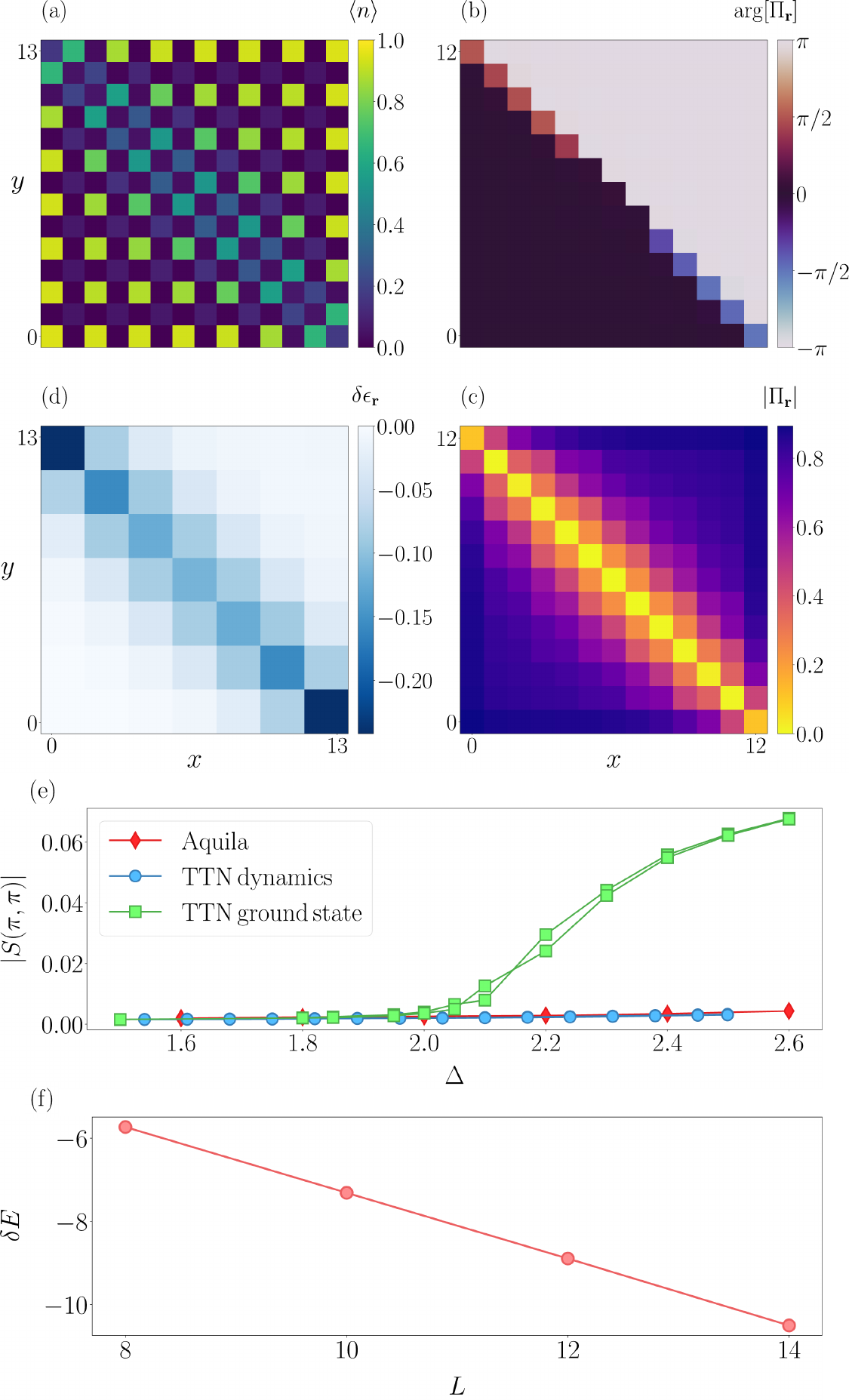}
    \caption{\justifying
Ground state $\mathcal{C}_2$ structure at $\Delta=2.5$.
(a) Local Rydberg density $\langle n_{\mathbf{r}}\rangle$, showing two striated domains nucleated by two opposite corners separated by a diagonal interface.
(b) Phase of the complex plaquette order parameter $\Pi_{\mathbf{r}}$.
(c) Modulus $|\Pi_{\mathbf{r}}|$, whose suppression along the diagonal marks the localized geometrical defect.
(d) Plaquette-averaged local energy density $\langle\delta\epsilon_{\mathbf{r}}\rangle$, showing that the energy gain is concentrated along the diagonal interface.
(e) The structure factor $S(\pi,\pi)$ computed as a function of $\Delta$. In the ground state (green squares), it increases from zero at $\Delta \simeq 2.0$, signaling the diagonal-defect regime. In the evolved state computed with TDVP simulations (blue circles) and in the experiment (red diamonds) it remains close to zero.
The two green curves correspond to simulations with different bond dimensions ($m=200,250$), with relative convergence in energy up to $10^{-3}$.
(f) The total energy difference $\delta E$ vs the lattice size $L$ for $\Delta = 2.5$. The modulus of $\delta E$ increases linearly in $L$, consistent with the support on which the gain is concentrated. 
}
\label{fig:FIG4_Z2_compl_pi_tot}
\end{figure}
This structure is shown in Fig.~\ref{fig:FIG4_Z2_compl_pi_tot}(a--c), where we plot the local density, and the phase and modulus of $\Pi_\mathbf{r}$ for one of the two $\mathcal{C}_2$-symmetric states, as obtained via DMRG, at $L=14$ and $\Delta=2.5$. 
In this regime, boundary-induced frustration produces a geometrical defect that no longer spreads from the center of the lattice, but is located along the lattice diagonal connecting the two unexcited corners.

The unexcited corners relieve the frustration along the lattice edges: with the domain walls along the boundaries removed, the lattice boundary forms a defect-free antiferromagnetic pattern. Similarly for the bulk: each $\mathcal{C}_2$ configuration contains two bulk domains in the same $\mathbb{Z}_2$ sector.
Each configuration thus independently develops a nonvanishing staggered magnetization.
In a symmetric superposition restoring the $\mathcal{C}_4$ rotational symmetry, however, the two configurations give opposite contributions, resulting in a vanishing total staggered magnetization.
To robustly characterize the diagonal-defect regime and detect the emerging $\mathbb{Z}_2$ ordering, we compute the $\mathbf{k}=(\pi,\pi)$ component of the static density-density structure factor $S(\mathbf{k})=1/L^4\sum_{\mathbf{r},\mathbf{r}'}\langle n_\mathbf{r}n_{\mathbf{r}'}\rangle\exp[-i\mathbf{k}\cdot(\mathbf{r}-\mathbf{r}')]$.

In Fig.~\ref{fig:FIG4_Z2_compl_pi_tot}(e), we show $|S(\pi,\pi)|$ as a function of $\Delta$ computed on the DMRG ground state, the TDVP one, and the experimental one. The structure factor of the DMRG ground state remains close to zero in the $\mathcal{C}_4$-symmetric frustrated regime, and increases at $\Delta\gtrsim2.0$, signaling the onset of long-range staggered order. 
To investigate the energetic origin of the frustrated states, we compare the energy density of a geometrically frustrated configuration with that of a defect-free configuration with a single domain. \simone{To estimate the energy density of a single striated domain of even size $L$, we numerically simulate the ground state of an odd-sized, $(L+1 )\times (L+1)$ system, where the striated order extends across the array without frustration.}
Then, we compute the local energy density
\begin{equation}\label{eq:energy_density}
    \epsilon_\mathbf{r}:
    =
    \frac{1}{2}\langle\sigma^x_\mathbf{r}\rangle
    -
    \Delta \langle n_\mathbf{r}\rangle
    +
    \frac{1}{2}
    \left(\frac{R_b}{a}\right)^6
    \sum_{\mathbf{r}'}
    \frac{\langle n_\mathbf{r}n_{\mathbf{r}'}\rangle}
    {|\mathbf{r}-\mathbf{r}'|^6}.
\end{equation}
We restrict $\mathbf{r}$ to the $L\times L$ sublattice $\mathcal{E}$ obtained by removing one boundary row and one boundary column from the odd-sized lattice, and apply the same restriction to the interaction sum over $\mathbf{r}'$. Consistently with the numerical simulations, we also truncate the the latter at distances $d\leq2\sqrt{2}\,a$. 
For each $\mathbf{r}\in\mathcal{E}$, we define the energy density difference $\delta \epsilon_\mathbf{r} = \epsilon_\mathbf{r}|_{\mathrm{even}}-\epsilon_\mathbf{r}|_{\mathrm{odd}}$, where $\epsilon_\mathbf{r}|_{\mathrm{even}}$ and $\epsilon_\mathbf{r}|_{\mathrm{odd}}$ are computed over the ground states with frustration (even size $L$) and without (odd size $L+1$).
In Fig.~\ref{fig:FIG4_Z2_compl_pi_tot}(d), we show the difference in the local energy $\delta\epsilon_\mathbf{r}$, coarse-grained over $2\times2$ plaquettes to remove oscillations associated with the striated pattern. 
\simone{We find that
$\delta\epsilon_\mathbf{r}$ is negative and localized along the same diagonal that separates the two domains, indicating that the defect itself provides the energy gain that makes the $\mathcal{C}_2$ configurations stable against single-domain states.
} 

\simone{The localization of the energy gain along the lattice diagonal allows us to address whether the regime shown here persists at large $L$. 
To this end, we compute the ground state at different lattice sizes $L=8,10,12,14$ at $\Delta=2.5$, where we find the same $\mathcal{C}_2$ configuration. 
Then, we compute the total energy difference $\delta E$ by integrating the difference in the local energy $\delta \epsilon_\mathbf{r}$. 
We find $\delta E<0$, with its modulus increasing linearly with $L$, as shown in Fig.~\ref{fig:FIG4_Z2_compl_pi_tot}(f). If the total energy gain $\delta E$ were spread over $\sim$$L^2$ bulk sites, it would vanish locally in the limit of large $L$. Instead, since the energy density difference $\delta \epsilon_\mathbf{r}$ concentrates along the diagonal, on a support scaling linearly with $L$, it remains finite locally as we increase $L$. 
Therefore, the diagonal defect remains stable against $L$, and we argue that the $\mathcal{C}_2$ configurations are robust at arbitrarily large even $L$. 
Even though at finite $L$, the ground state contains the superposition of the two $\mathcal{C}_2$ configurations, remaining $\mathcal{C}_4$-symmetric, variational simulations tend to select one of the two as the energy splitting decreases with the system size. 
While at $L=8$ we capture the $\mathcal{C}_4$-symmetric superposition of the two $\mathcal{C}_2$ configurations, the numerical ground state selects one of them at larger $L$ (see App.~\ref{app:C2symmbreak} for details). 
\simone{In the (even $L$) thermodynamic limit,} it is reasonable to expect that
the two $\mathcal{C}_2$ configurations become
degenerate ground states, leading to a spontaneous breaking of $\mathcal{C}_4$ symmetry. 
}

\simone{We finally address the detuning ramp protocol, as we cross $\Delta\simeq2.0$. As Fig.~\ref{fig:FIG4_Z2_compl_pi_tot} shows, in this case $|S(\pi,\pi)|$ is negligible compared to the corresponding DMRG ground state value.
This rules out the presence of a $\mathcal{C}_2$-symmetric component in the evolved state,
for both the TDVP simulations and the experiment.
To understand why the ramp remains trapped in the four-domains manifold, we observe that the dynamical formation of a diagonal-defect configuration would require the boundary domain walls to localize in pairs at two opposite lattice corners. 
If we model each domain wall as a particle freely moving along a lattice edge and
neglect the bulk dynamics, the probability of such a localization scales
polynomially with $L$. Since each boundary domain wall also selects a set of compatible
bulk configurations, we expect this probability to be further suppressed.
Furthermore, boundary frustration likely generates a dense set of low-lying $\mathcal{C}_4$-symmetric states populated during the dynamics.
} 

\section{discussion and conclusions}
\label{sec:conclusions}

In this work, we show that boundaries can act as a source of frustration in two-dimensional Rydberg atom arrays, pinning the bulk ordering pattern, rather than merely perturbing it.
Building on past results that prepared the striated phase in a single ordered domain~\cite{Ebadi2021QuantumPhases} and pointed out distinct bulk-boundary critical behaviors~\cite{bulk_boundary_kalinowski_2022}, we compare the cases where the lattice size is commensurate or not with the striated pattern. Our main result is a clear distinction in the low-energy sector between odd and even arrays emerging in the striated density-wave phase.
The experiments on the Aquila quantum processor, together with tree tensor network simulations, show that the lattice parity acts as a geometrical constraint that reshapes the ordering structure, despite identical Hamiltonian couplings.

At equilibrium, we have shown two different regimes at even sizes that have no counterpart at odd sizes. 
At detuning $\Delta \lesssim 2.0$, the state obtained from ground-state numerical simulations contains four incompatible domains seeded by the lattice corners.
Using a local order parameter that distinguishes the four domains, we observed in the experiment and in real-time simulations a central defect emanating from a point singularity in the phase of the order parameter, around which the phase of $\Pi_\mathbf{r}$ winds by $2\pi$.
At $\Delta \gtrsim 2.0$, the ground state contains two domains seeded by opposite corners, separated by a one-dimensional diagonal defect. 
Although the ground state at finite size remains $\mathcal{C}_4$-symmetric, a new $\mathcal{C}_2$ order emerges, represented by a $\mathbb{Z}_2$, Ising-like antiferromagnet. We argue that the $\mathcal{C}_4$ symmetry breaks spontaneously down to $\mathcal{C}_2$ in the limit of large even $L$.

The experiment and out-of-equilibrium simulations enrich this picture. We find that during a linear ramp of the detuning, the state enters the $\mathcal{C}_4$ regime, as the corners seed the domains and the boundary striated order grows in time. Yet, as we increase the detuning further, the same domains act as geometrical constraints that prevent the state from entering the $\mathcal{C}_2$ regime. We argue that a set of low-lying states with four excited corners traps the state in the $\mathcal{C}_4$ regime. Preparing the $\mathcal{C}_2$ configuration could be facilitated by local control tools, such as local detunings, available on Aquila, extending previous work on one- and two-dimensional systems~\cite{Omran2019CatStates,manovitz_quantum_2025} and irregular lattices~\cite{bombieri_PRXQuantum.6.020306}. Probing the $\mathcal{C}_2$ regime would allow us to investigate equilibrium and out-of-equilibrium properties of the observed defect, and constitutes a natural outlook for our work.

Another interesting question is how the observed phenomenology depends on the system size. 
We show that $\mathcal{C}_2$ configurations are stable against single-domain states by an energy gain localized along their diagonal interface.
Since this energy gain grows linearly in $L$ on a spatial support linear in $L$, we expect the $\mathcal{C}_2$ regime to emerge at every even $L$. 

Although the phenomena we show here are limited to a precise Hamiltonian parameter regime, they show that boundary effects can robustly influence bulk properties beyond one dimension.
This raises the question of the role played by the lattice geometry and the long-range interactions. 
Moreover, it might be possible to employ the boundaries to engineer geometrical defects with different codimensions, which are expected to differ in their mobility and stability under a drive.

\begin{acknowledgments}
We thank Milan Kornja{\v{c}}a and Arinjoy De for their feedback on the manuscript.
We thank Tommaso Macr\`i, Alan Bidart, Phillip Weinberg for assistance in the use of the Aquila quantum simulator.
SN thanks Jonathan Wurtz, Sheng-Tao Wang, and Fangli Liu for hosting and for fruitful discussions at QuEra Computing.
SN acknowledges support from Mikhail Lukin and the Harvard Physics Department  during the development of this project.

TTN calculations were performed with the Quantum TEA library~\cite{qtealeaves,Jaschke_2026}. We thank Oliver Bellwood for providing the mapping presented in Ref.~\cite{Bellwood2025} for the $L=14$ lattice. 
We acknowledge computational resources from the INFN Padova
HPC cluster, funded by the NextGenerationEU Terabit project
on PNRR – Avviso n.~3264 ``per il Rafforzamento e creazione
di Infrastrutture di Ricerca'', Missione 4, ``Istruzione e
Ricerca'', and from the Leonardo supercomputer hosted by
CINECA (Italy) and the LEONARDO consortium.
This work was supported in part by the high-performance computing infrastructure developed under the project ``CONVECS'', funded by the PR Veneto FESR 2021-2027 program, Priority 1 – Specific Objective 1.1 – Action 1.1.2

This work received support from IMPAQT DARPA-EA-23-01-01 Agreement No.~HR0011-23-3-0018, from the European Union via
ICSC – Italian Research Center on HPC, Big Data and
Quantum Computing (NextGenerationEU Project No.~CN00000013), Horizon Europe program HORIZON-CL4-
2022-QUANTUM-02-SGA via the project 101113690
(PASQuanS2.1), by the Italian Ministry of University and
Research (MUR) via Quantum Frontiers (the Departments
of Excellence 2023-2027), by the World Class Research
Infrastructure – Quantum Computing and Simulation Center
(QCSC) of Padova University, and by the Istituto Nazionale di
Fisica Nucleare (INFN): iniziativa specifica IS-QUANTUM. 

SN acknowledges that this project has received funding from
the European Union’s Horizon Europe research and innovation program under the Marie Sklodowska-Curie grant
agreement No.~101059826 (ETNA4Ryd). 
Work at MPQ is part of the Munich Quantum Valley, which is supported by the Bavarian state government with funds from the Hightech Agenda Bayern Plus.

\end{acknowledgments}
\section*{Data availability}
Numerical and experimental data are available upon reasonable request.
\bibliography{arxiv_bibliography}%
\appendix

\section{Tree tensor networks}
\label{app:ttns}

The numerical results shown in this work rely on tensor-network methods. We use the Tree-Tensor-Network (TTN)~\cite{felser_introduction_2026} Ansatz to represent the quantum many-body state. In a TTN, tensors are arranged into a loopless tree structure. The tensors in the bottom layer carry the physical degrees of freedom, while the upper layers act as auxiliary tensors that efficiently mediate correlations across the network.

The absence of loops is a key ingredient for the efficiency of TTNs, since it allows observables and correlation functions to be contracted efficiently. Moreover, many algorithms originally developed for Matrix-Product States can be generalized to tree geometries. We use the time-dependent variational principle (TDVP)~\cite{Haegeman2016,Bauernfeind_2020} for real-time evolution. In particular, we adopt the single-site version of TDVP, where one tensor is updated at a time. Since this algorithm does not increase the bond dimension during the evolution, we initialize the state already at the maximal bond dimension used in the simulation. 

A two-dimensional lattice must be embedded into the tree geometry through a mapping of the physical sites onto the physical legs of the network. This mapping can influence the accuracy of the simulation, as physically local couplings in the two-dimensional lattice become long-range in the network.
This generates correlations between distant branches of the tree and requires, in principle, a larger bond dimension to accurately represent the state. The mapping must be chosen such that it minimizes the number and length of these effective long-range connections. 
Binary tree tensor networks are naturally suited to represent systems with the size of a power of two, for which the Hilbert curve~\cite{Cataldi_2021} is a natural choice. 
Otherwise, we embed the system into the smallest binary tree containing the system, and ignore the remaining redundant tensors. For $8<L\leq16$, we use $N_{tot}=256$ physical legs.
We use the standard snake map for $L=9,10,11,12,13,15$, while for $L=14$ we use an optimal self-similar mapping introduced in Ref.~\cite{Bellwood2025}, which improved the ground-state search. For $L=8,16$ we use the standard Hilbert mapping~\cite{Cataldi_2021}.

\begin{figure}
\centering
    \includegraphics[width=0.5\linewidth]{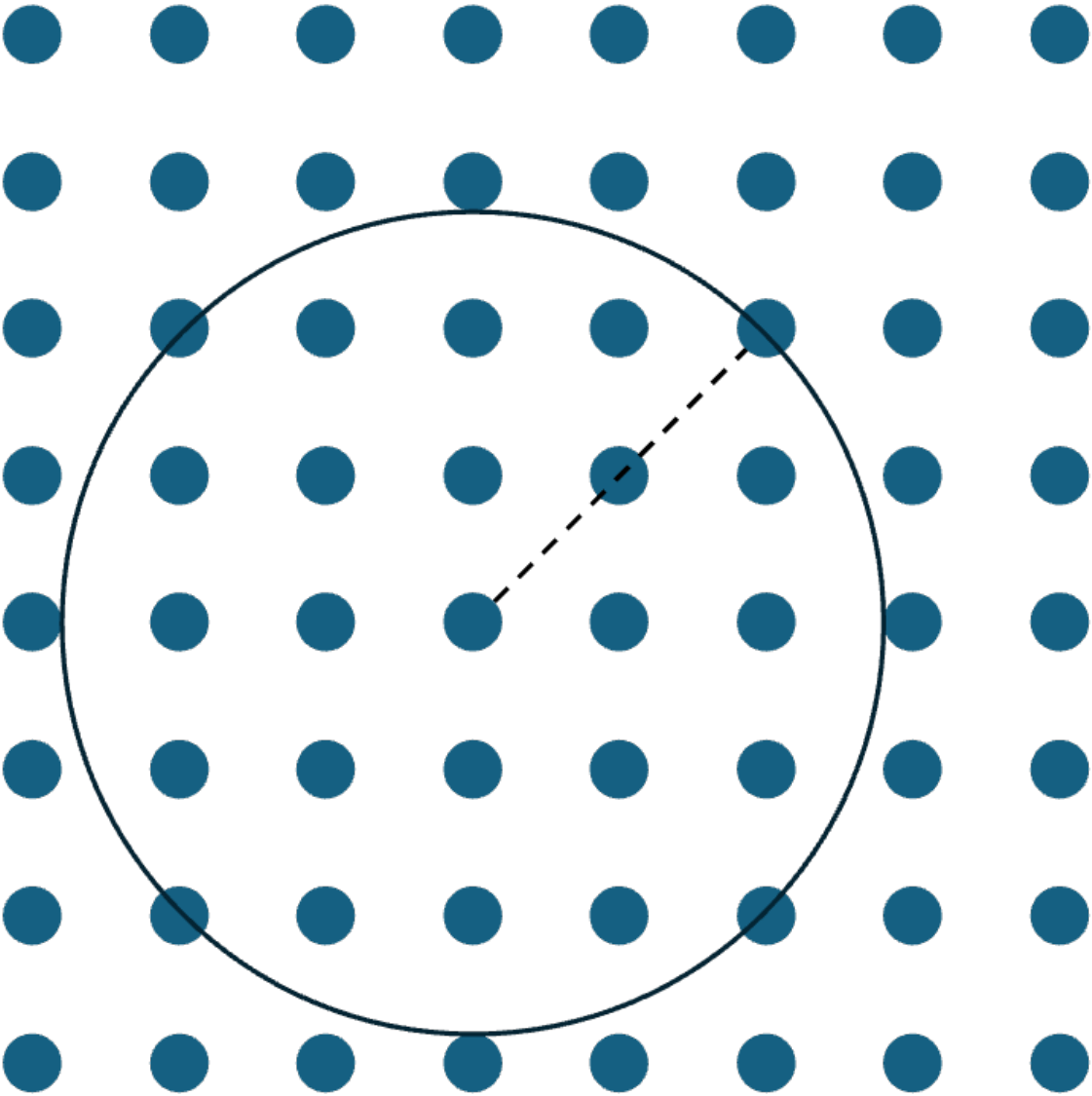}
    \caption{\justifying
Sketch of the interaction range we retain in numerical simulations.
}
\label{fig:app_truncation}
\end{figure}

In the definition of the model Hamiltonian, we truncate long-range interactions up to $d_{\rm max} = 2\sqrt{2}a$ as defined in the main text, which corresponds to the radius shown in Fig.\ref{fig:app_truncation}. The corresponding energy is $V(d_\mathrm{max}) = V_1/512$.

For the real-time simulations, we simulate the whole experimental protocol, starting from an initial state with all the atoms in their ground state. Nevertheless, the relevant part for the analysis is the linear ramp across the critical region, the white one in Fig.~\ref{fig:FIG1}(e).

For ground-state density matrix renormalization group (DMRG) numerical simulations, we initialize a random TTN and iteratively optimize its tensors to minimize the energy of the state. We repeat the sweep optimization up to a maximum number of 150 sweeps.
For all dynamic simulations, we use the same time step. Using the convention explained in App.~\ref{app:experimental_details}, we set $d\tau=0.002$, smaller than $0.01/\Delta$ in the relevant parameter regime. The relevant energy scale controlling the dynamics is set by the detuning rather than by the nearest-neighbor interaction strength, since states violating the blockade constraint are energetically suppressed and affect the low-energy dynamics only perturbatively.

\section{Disordered-to-striated phase transition with odd \texorpdfstring{$L$}{L}}
\label{app:extr_QPT_odd}

\begin{figure}
\centering
    \includegraphics[width=1.0\linewidth]{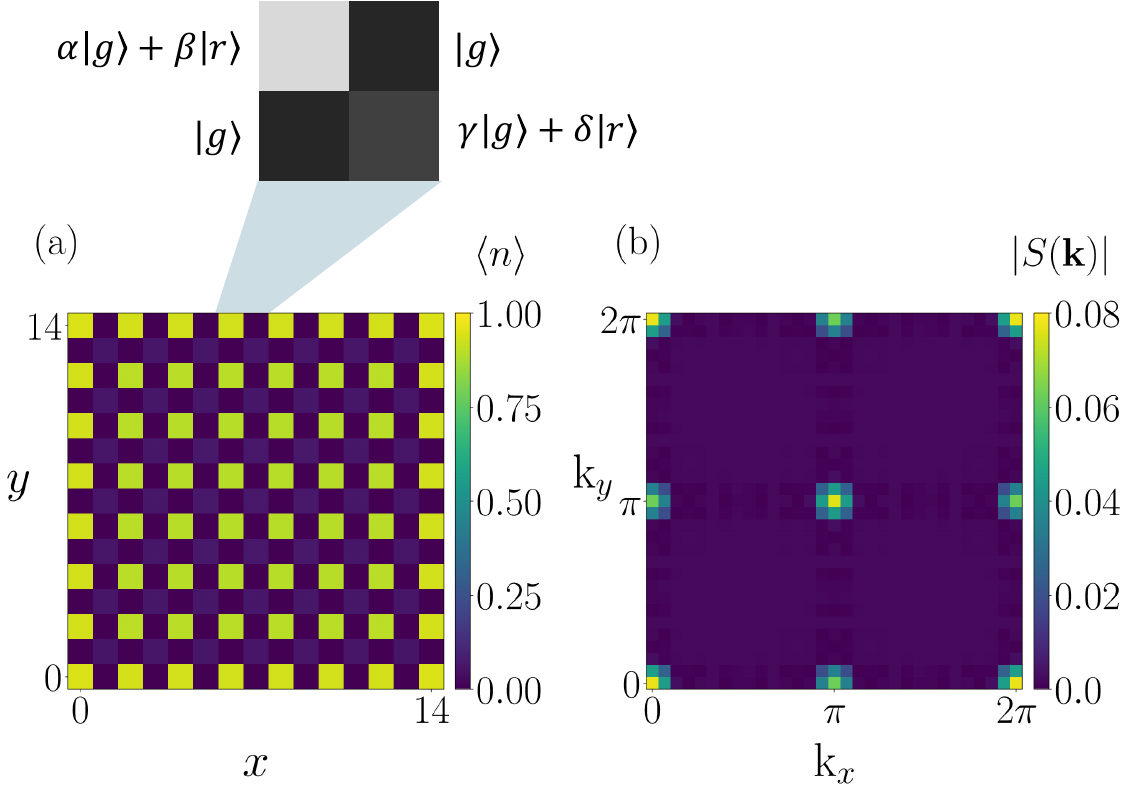}
    \caption{\justifying
(a) Density profile in the striated phase computed on a $15\times 15$ lattice at $\Delta=2.5$. The order over a $2\times2$ plaquette is sketched above, with two sites in their ground state and the other two in a superposition such that $\beta \simeq \gamma\simeq1$ and $\alpha\simeq\delta\ll1$.
(b) The striated order is captured by the $(\pi,0)$ and $(0,\pi)$ components of the structure factor.
 }
\label{fig:striated_fig}
\end{figure}

The striated phase is a quantum phase characterized by a four-fold degenerate density pattern defined over a $2\times 2$ plaquette: Two diagonally opposite sites are approximately in their ground state $|g\rangle$, while the other two are in a strongly imbalanced ground -- Rydberg state $|r\rangle$ (see the sketch in Fig.~\ref{fig:striated_fig}(a)).
Importantly, the classical configuration with three sites in their ground state and one excited would not be energetically stable. In fact, the phase is stabilized by the weak excitation of the bottom-right site in the plaquette ($\delta\ll1$), due to quantum fluctuations~\cite{Ebadi2021QuantumPhases}. This pattern breaks the horizontal and vertical translational invariances separately, and it is captured by the $(0,\pi)$ component of the structure factor $S(\mathbf{k})=1/L^4\sum_{\mathbf{r},\mathbf{r}'}\langle n_\mathbf{r}n_{\mathbf{r}'}\rangle\exp[-i\mathbf{k}\cdot(\mathbf{r}-\mathbf{r}')]$, while the $(\pi,\pi)$ component captures the imbalance between the two checkerboard sublattices (see Fig.~\ref{fig:striated_fig}(b)).

To study the disordered-to-striated phase transition, we set $R_b/a=1.46$.
At values of $\Delta \gtrsim \Omega$, the boundary orders first~\cite{bulk_boundary_kalinowski_2022}, while the bulk enters the spin-density-wave striated phase at a larger detuning~\cite{Ebadi2021QuantumPhases}. It has been shown~\cite{bulk_boundary_kalinowski_2022} that the boundary exhibits a second-order, (1+1)-dimensional Ising quantum phase transition between the disordered phase and an antiferromagnetic phase. When $L$ is odd, 
the boundary forms a defectless configuration with the four corners excited, as it minimizes the interaction energy. 

\begin{figure}
\centering
    \includegraphics[width=1.0\linewidth]{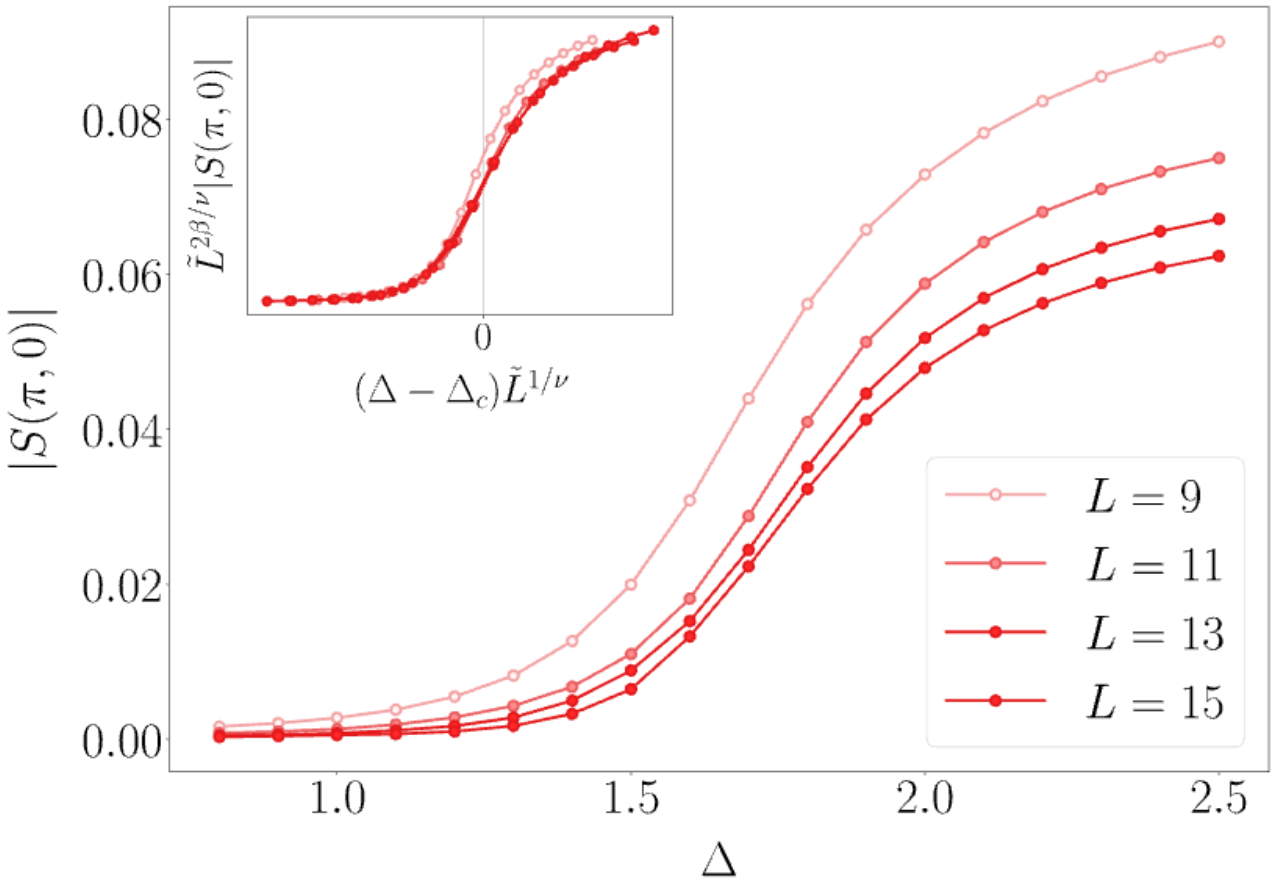}
    \caption{\justifying
Finite-size scaling analysis of the disordered-to-striated transition in odd-sized square arrays. 
Main panel: $(\pi,0)$ component of the static density-density structure factor, $|S(\pi,0)|$, computed from TTN ground-state simulations for odd lattice sizes $L=9,11,13,15$ as a function of the detuning $\Delta$. For the finite-size-scaling procedure, we excluded $L=9$ to obtain the best collapse. The structure factor is taken in the bulk of the lattice, so the effective length of the system is $\tilde L = L-2$.
Inset: the best collapse of the rescaled structure factor $\tilde L^{2\beta/\nu}|S(\pi,0)|$ as a function of the scaling variable $(\Delta-\Delta_c)\tilde L^{1/\nu}$.
}
\label{fig:odd_finite_size_scaling}
\end{figure}

It has been experimentally observed~\cite{Ebadi2021QuantumPhases} that the boundary configuration pins a striated configuration in the bulk, facilitating the observation of the striated phase. 
In particular, the system orders first along the boundary and then in the bulk as the detuning is increased, consistent with a so-called extraordinary phase transition~\cite{Diehl1997Boundary,ParisenToldin2022Boundary,Padayasi2022Extraordinary}. 
We show here a scaling behavior of the striated order parameter, which is compatible with a second-order phase transition as well. 
We compute $S(\pi,0)$ on the ground state of lattices with $L=9,11,13,15$, restricting the sum to bulk sites and excluding autocorrelators $\langle  n_\mathbf{r}  n_\mathbf{r}\rangle$, so that the effective linear size is $\tilde L=L-2$. The
maximum of $\mathrm{d}S/\mathrm{d}\Delta$ is stable at $\Delta\simeq1.75$ for
$L=11,13,15$, which we take as an estimate of $\Delta_c$, while for $L=9$ it deviates
significantly, signaling strong finite-size corrections; we therefore exclude $L=9$ from
the collapse. Fixing $\Delta_c=1.75$, the scaling of $S(\pi,0)$ at $\Delta_c$ yields
$2\beta/\nu\simeq0.7$, while $\nu$ is poorly constrained by our system sizes
($\nu\simeq0.9$--$2.2$, depending on the fitting window). The inset of Fig.~\ref{fig:odd_finite_size_scaling}
shows the collapse for $2\beta/\nu=0.67$ and $\nu=1.06$.
We compute the ground state of an $L\times L$ lattice with $L=9, 11, 13, 15$. From the density-density correlators $\langle  n_\mathbf{r}  n_\mathbf{r'}\rangle$, we compute the static structure factor  $S(\mathbf{k})$ with $\mathbf{k} = (\pi,0)$. 

Although the scaling relation agrees with a second-order phase transition, the limited set of sizes we consider prevents us from inferring the universality class. Nevertheless, the critical point $\Delta_c \simeq 1.75$ is larger than the critical value of the boundary transition found in Ref.~\cite{bulk_boundary_kalinowski_2022}, consistent with an extraordinary phase transition, and confirming the bulk-boundary interplay in the system.

\section{Experimental protocol and data post-selection}
\label{app:experimental_details}

\begin{table}
\caption{\justifying Number of shots for each final detuning $\Delta$ ($L=14$): total, retained after the filling post-selection (at most one missing atom), and after the additional blockade post-selection (at most one nearest-neighbor violation). The last column gives the mean blockade-violation probability per site $p_{\mathrm{bv}}$ obtained from the ratio of the last two shot counts by modeling violations as Poisson-distributed independent events.}\label{tab:shots}
\begin{ruledtabular}
\begin{tabular}{ccccc}
$\Delta$ & Total & Filling & Filling + blockade & $p_{\mathrm{bv}}$ (\%) \\
\hline
1.0 & 4800 & 3296 & 2812 & 0.34 \\
1.2 & 2400 & 1793 & 1549 & 0.33 \\
1.4 & 4400 & 2735 & 2124 & 0.45 \\
1.6 & 2400 & 1763 & 1444 & 0.39 \\
1.8 & 2400 & 1751 & 1407 & 0.42 \\
2.0 & 4400 & 2738 & 1919 & 0.56 \\
2.2 & 2400 & 1755 & 1315 & 0.49 \\
2.4 & 3400 & 2076 & 1444 & 0.57 \\
2.6 & 2400 & 1740 & 1268 & 0.52 \\
\end{tabular}
\end{ruledtabular}
\end{table}

Experiments are performed on the cloud-accessible Aquila neutral-atom quantum processor developed by QuEra. We use square arrays of ${}^{87}$Rb atoms with lattice spacing $a$, and choose the Rydberg blockade radius such that $R_b/a=1.46$. The system is driven by the global Rydberg Hamiltonian introduced in the main text. All energy scales are expressed in units of the maximum Rabi frequency allowed by the device $\Omega_{\max}$. In these units, the nearest-neighbor interaction is $V_1=C_6/(a^6\Omega_{\max})=(R_b/a)^6$.

The preparation protocol consists of an initial turn-on of the Rabi drive followed by a detuning ramp. The detuning is initialized at $\Delta_i=-2.0$, in units of $\Omega_{\max}$, and is first brought to $\Delta_0=0$. We then linearly ramp the detuning from $\Delta_0$ to the final value $\Delta_f$, which is denoted by $\Delta$ throughout the main text. At the end of the ramp, the Rabi drive is turned off as part of the experimental readout sequence.
In the real-time tensor-network simulations, we exclude the final turn-off of the Rabi drive, present in the experimental sequence.

We measure time in units of $2\pi/\Omega_{\max} \approx 0.40\,\mu\mathrm{s}$,
$\tau = t\,\Omega_{\max}/2\pi$, and energies in units of $\Omega_{\max}$, so that
the state evolves as $i\,\partial_\tau |\psi\rangle = 2\pi {H}(\tau)\,|\psi\rangle$.
In these units, the linear detuning ramp reads $\Delta(\tau) = v_\Delta(\tau - \tau_0)$,
where $\tau_0$ marks the start of the ramp and $v_\Delta = 0.7$ is the ramp rate,
corresponding to $\mathrm{d}\Delta/\mathrm{d}t = v_\Delta\,\Omega_{\max}^2/2\pi
\approx 27.8\,\mathrm{rad}/\mu\mathrm{s}^2$.
The ramp from $\Delta = 0$ to $\Delta_f$ thus lasts $T_\Delta = \Delta_f/v_\Delta$,
e.g.\ $T_\Delta \approx 2.86$ ($\approx 1.14\,\mu\mathrm{s}$) for $\Delta_f = 2$.

Each experimental repetition yields a projective measurement of the Rydberg excitation pattern. Before computing densities and correlation functions, we apply a post-selection procedure to remove shots affected by preparation defects (missing atoms) or clear violations of the blockade constraint (excitations closer than the blockade radius).  We retain shots containing at most one initial lattice defect, and we assign the value $0$ to the corresponding site, indicating an unexcited atom. At the end of the protocol, we retain shots containing at most one nearest-neighbor blockade violation, provided that the violating cluster consists of a single nearest-neighbor pair. Although closer than the blockade radius, we do not exclude diagonal-neighboring excitations, as they characterize the striated phase. 
The remaining shots are used to estimate the local Rydberg density, plaquette order parameter, and spatial density-density correlations reported in the main text.

For each value of $\Delta$, we obtained around $1300$--$2800$ shots after initial filling and blockade-violation post-selection. Assuming independent vacancies, we estimate an average initial-filling probability of $\sim 99.4\%$. 
Then, we observe that the probability of nearest-neighbor blockade violations grows with the
total protocol duration, which increases linearly with $\Delta$, with an averaged per-site blockade violation probability of $\sim 0.45\%$. We conjecture that the observed increase is due to, for example, atoms
lost during the evolution, as they are read out as Rydberg excitations. We report the relevant data in Table~\ref{tab:shots}.

Error bars are estimated via bootstrap standard deviations over experimental shots and are smaller than the marker size where not visible.

\section{Variational symmetry selection in ground-state simulations}
\label{app:gs_symmetry_selection}

In the even-sized frustrated regime, the low-energy manifold contains configurations in which the boundary-induced domains and the associated bulk defect lines are localized in different positions. In the exact finite-size ground state, lattice symmetries imply that symmetry-related localized configurations should be combined into a globally symmetric superposition. In practice, however, variational ground-state tensor-network simulations at finite bond dimension may converge to one representative, or to a restricted subset, of these localized configurations.

This effect is analogous to the spontaneous symmetry breaking commonly observed in variational simulations of finite systems with nearly degenerate symmetry-related states. When the degeneracy of different localized domain-wall configurations is high, and their energy difference is small, the variational mapping and finite bond dimension can favor a lower-entanglement representative of the manifold, thereby making the symmetry breaking explicit at the level of local observables. As a result, the computed ground-state density and plaquette-order profiles may display domains pinned in asymmetric positions, even though the exact ground state would restore the symmetry through a superposition of symmetry-related configurations.

This variational symmetry selection affects observables that are sensitive to the spatial averaging over domain-wall positions. In particular, the rotational correlator $\Phi$ can be enhanced when the variational state is localized around configurations in which the bulk defect lines intersect coherently, because the configurational averaging over different domain-wall positions is reduced. Therefore, the value of $\Phi$ extracted from ground-state tensor-network calculations should be interpreted as probing a representative localized component of the low-energy manifold, rather than necessarily the fully symmetry-restored finite-size ground state.

By contrast, this issue is less pronounced in the real-time simulations of the experimental ramp. The time evolution is initialized from a symmetric state and follows a symmetric Hamiltonian. In the absence of explicit symmetry-breaking perturbations, the dynamics preserves the global lattice symmetries. Moreover, the correlations between boundary domains are built progressively during the ramp and remain encoded in the evolved state. Consequently, the real-time tensor-network simulations maintain the globally $\mathcal{C}_4$-symmetric structure observed experimentally, even in parameter regimes where ground-state optimization can converge to localized configurations.

This distinction provides a natural explanation for the different behavior of the rotational correlator in ground-state and dynamical simulations. Ground-state calculations reveal the structure of the low-energy manifold and its tendency toward localized defect configurations, while the ramp dynamics probes the symmetry sector dynamically accessed from the initial state. The comparison between the two therefore supports the interpretation of the frustrated regime as a dense manifold of low-lying boundary-defect configurations, whose symmetry properties depend on both the preparation protocol and the variational representation used to approximate the state.

\section{$\mathcal{C}_2$ regime and symmetry breaking}
\label{app:C2symmbreak}
\begin{figure}[t]
\centering
    \includegraphics[width=1.0\linewidth]{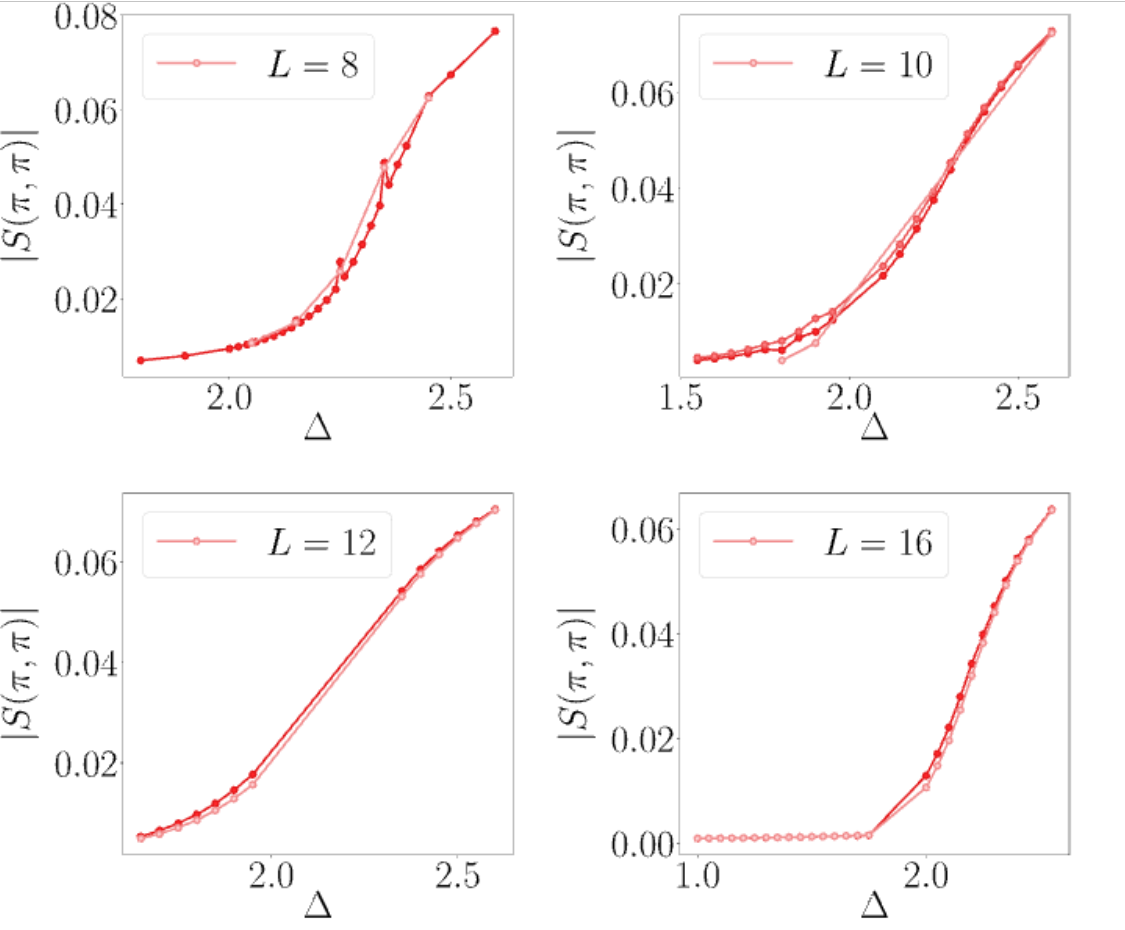}
    \caption{\justifying
Structure factor $|S(\pi, \pi)|$ at even $L=8,\dots,16$. Different shades show results obtained at bond dimension $m=200,250$ with relative energy convergence at $10^{-3}$.
}
\label{fig:FIG9}
\end{figure}

Here, we discuss the numerical results we obtain at different lattice sizes in the $\mathcal{C}_2$ regime. 
In Fig.~\ref{fig:FIG9}, we first show that the $\mathcal{C}_2$ regime manifests at $L=8,\dots, 16$. 
\begin{figure*}
\centering
    \includegraphics[width=1.0\linewidth]{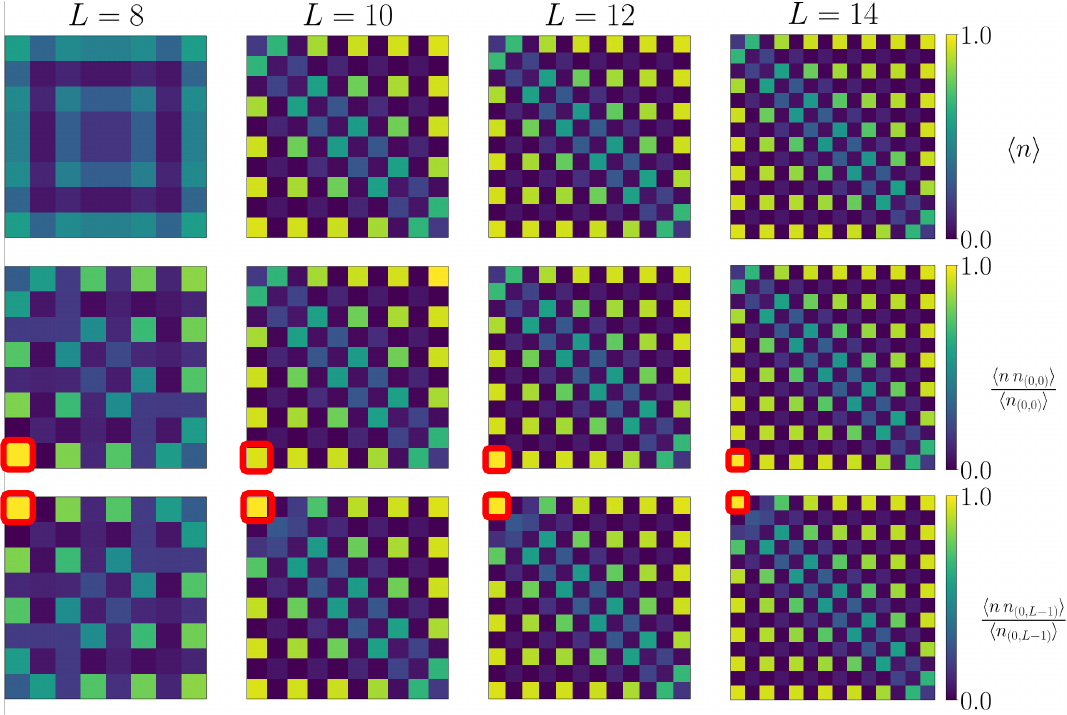}
    \caption{\justifying
Analysis of the $\mathcal{C}_2$ ground state obtained from numerical simulations.
First row: Average Rydberg excitation density. Second row:  Density-density correlator between the bottom-left corner and the rest of the lattice.
Third row: Density-density correlator between the top-left corner and the rest of the lattice. Correlators are rescaled with the density of the respective fixed points, circled in red.}
\label{fig:app_even_L_C2}
\end{figure*}
Then, we compare the results we obtain for $L=8,10,12,14$ that we use to produce Fig.~\ref{fig:FIG4_Z2_compl_pi_tot}(f). In the first row of Fig.~\ref{fig:app_even_L_C2}, we plot the density $\langle n \rangle$ at different sizes. We notice that, for $L=10,12,14$, the simulation selects a single $\mathcal{C}_2$ configuration. Instead, for $L=8$, the density profile exhibits $\mathcal{C}_4$ invariance. To investigate this behavior, we consider the density-density correlator in which we fix the top-left (second row) and bottom-left (third row) corners, respectively. For $L=8$, we recover the $\mathcal{C}_2$ configurations aligned along the two diagonals. This means the simulation captures the superposition of the two, yielding a $\mathcal{C}_4$-invariant state. At larger sizes, the simulation instead collapses into a single $\mathcal{C}_2$ configuration, artificially breaking the $\mathcal{C}_4$ invariance of the ground state. For example, at $L=10$, when plotting the correlator with the bottom-left corner fixed, the pattern we observe contains the same diagonal defect. Unlike $L=8$, the correlator with the top-left corner fixed does not show the opposite $\mathcal{C}_2$ pattern, which would imply long-range correlation and require a higher bond dimension. Instead, we observe a similar pattern with the same diagonal, in which the top-left excited corner acts as a local defect.

\end{document}